# Gelation effects on the spreading of non-Newtonian drops impacting a reactive liquid

Julie Godefroid[1-2], David Bouttes[2], Anselmo Pereira[3,*] & Cécile Monteux[1,**]

[1]Sciences et Ingénierie de la Matière Molle (SIMM), ESPCI Paris, Université PSL, UMR CNRS 7615, Sorbonne Université, Paris, France

[2]Saint-Gobain CREE, France

[3]Mines Paris, Université PSL, Centre for Material Forming (Cemef), UMR CNRS 7635, rue Claude Daunesse, 06904 Sophia-Antipolis, France

**Abstract**: We report in this experimental and numerical study effects of gelation on the early-time spreading (< 10ms) of millimetric non-Newtonian drops of biopolymer and particle suspensions impacting a Newtonian liquid containing reactive compounds. Our analyses are initially conducted through experiments by considering a variety of biopolymer and gelling compound concentrations, and impact velocities. The experimental results are then compared to three-dimensional numerical simulations based on a variational multiscale approach, which focuses on the impact of simple viscoplastic drops on a liquid bath in the absence of gelation (non-reactive liquid). These comparisons enable the development of theoretical arguments that emphasise relevant gelation effects on the drop's maximum spreading.

## 1. Introduction

A challenging problem that couples interfacial hydrodynamics, rheology, and material phase change concerns the impact of non-Newtonian drops on a chemically reactive Newtonian liquid, inducing drop gelation [1]. Specifically, the drop, initially a weak gel, solidifies as it penetrates the liquid, becoming a strong gel [2-8]. This fundamental and scarcely explored physicochemical hydrodynamics topic [9] is directly related to essential health and industrial applications, such as encapsulation processes (for the protection of transplanted pancreatic islets cells, for example), 3D-bioprinting of cells, tissues, organs, prosthetics, and wound dressings; and ceramic beads and non-spherical solid particles production [10-14].

Typically, as illustrated by the experimental snapshots in Fig.1, a falling non-Newtonian polymer-based millimetric drop (underlined by the black rectangle) impacts a liquid/air interface, spreads (red rectangle), and forms an air cavity, which later retracts under capillary and gravity effects, while the drop penetrates the liquid containing reactive compounds (blue rectangle). The latter diffuses inside the drop, cross-linking the macromolecules and subsequently forming a gelled elastic membrane in the outer part of the drop. Over time, the impacting object transforms into a strong gel, a solid-like material [1-8, 15-20], while its shape relaxes into a spherical form (green rectangle). The morphological evolution of the drop is depicted at the bottom part of the figure (the weak gel appears in grey, and the elastic membrane is illustrated in shades of red). In short, during the reactive liquid entry, the drop undergoes both chemical-induced solidification (gelation) and deformation (impact, spreading and detachment from the air cavity), which, together with the shape relaxation induced at late times, leads to different final solid-like forms depending on the impact velocity, initial rheological properties of the drop, drop/liquid hydrodynamic interactions, and the gelation kinetics (polymers and gelling agent concentrations). Nevertheless, since non-Newtonian fluids can present an extremely rich rheological behaviour [which may include shear-rate-dependent viscosity, yield-stress, normal stress differences, elasticity, shear localization, etc.; 15-21], and the gelation kinetics is poorly understood, as well as its effects on the drop's rheology time evolution, the mentioned final shapes are difficult to predict and control. On a larger scale, the absence of such knowledge makes the development of the aforementioned applications complex. Consequently, there has been a growing industrial interest in fundamental studies on gelling non-Newtonian drops impacting liquids [1-14].

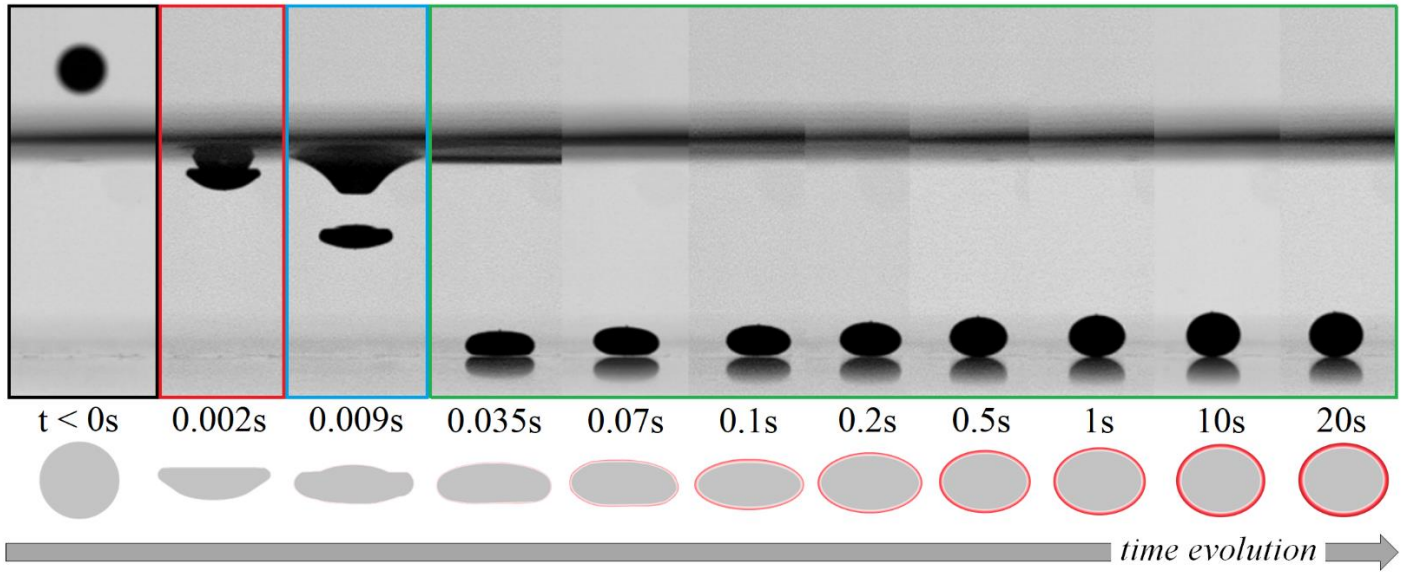

**Figure 1.** Experimental snapshots illustrating the penetration of an 8% by volume zirconia-in-deionised-water drop with 0.8% by volume of sodium alginate (biopolymer) into a liquid bath containing 158g/L of $CaCl_2 6H_2O$ (reactive compound). The falling 2.5mm-diameter non-Newtonian drop (in the black rectangle) impacts at the liquid/air interface at 1.4m/s, spreads (red rectangle), and forms an air cavity, which later retracts by capillary and gravitational effects (blue rectangle), while the drop penetrates the reactive liquid. As schematised at the bottom part of the figure, the reactive compounds diffuse inside the drop (weak gel represented in grey), cross-linking the macromolecules and consequently giving rise to a gelled elastic membrane in the outer part of the drop (strong gel illustrated in shades of red), whose shape relaxes towards a spherical form (green rectangle). A movie showing this flow case is available on https://www.youtube.com/watch?v=W48lB-GQUyQ.

Recently, the shape relaxation of elongated drops of biopolymer and particle suspensions under gelation (green rectangle in Fig.1), after their impact on a reactive liquid, was analysed [1]. It has been demonstrated through a combined experimental and numerical approach that the elastic shell formed by the crosslinking of polymeric macromolecules induces tensile stresses and drives a flow in the ungelled liquid core, resulting in the relaxation of the drops into spherical shapes. The thickness of this elastic membrane increases with time, allowing the bending stiffness required to change the drop's shape to eventually balance the elastic stresses within the elastic shell and halt the relaxation process. Despite the relevance of such a pioneering study, it is essential to note that it has exclusively considered gelation effects occurring at late times ($\gtrsim$ 10ms). Nevertheless, one can naturally wonder whether gelation affects the drop from the very beginning of the penetration process ($\lesssim$ 10ms). Hence, it is reasonable to extend the investigations to highlight potential gelation effects on the drop spreading at early times, immediately after its impact on the reactive liquid.

In the present work, we report, to the best of our knowledge, for the first time, relevant gelation effects on both the spreading and the morphological evolution of millimetric alginate-based drops impacting $CaCl_2 6H_2O$-in-deionised water baths. Our analyses are initially conducted through experiments using a variety of alginate and $CaCl_2 6H_2O$ concentrations. The experimental results are then compared to three-dimensional numerical simulations focused on the impact of simple viscoplastic drops on a water bath in the absence of gelation (non-reactive liquid). These comparisons enable the development of theoretical arguments that stress the important gelation effects on the drop's maximum spreading.

The organization of the paper is as follows. A detailed description of the physical formulation, and the experimental and numerical methods are presented in Section 2. Experimental and numerical results are discussed in Section 3. Finally, conclusions and perspectives are drawn in the closing section.

Email address for correspondence: * anselmo.soeiro pereira@minesparis.psl.eu | ** cecile.monteux@espci.fr

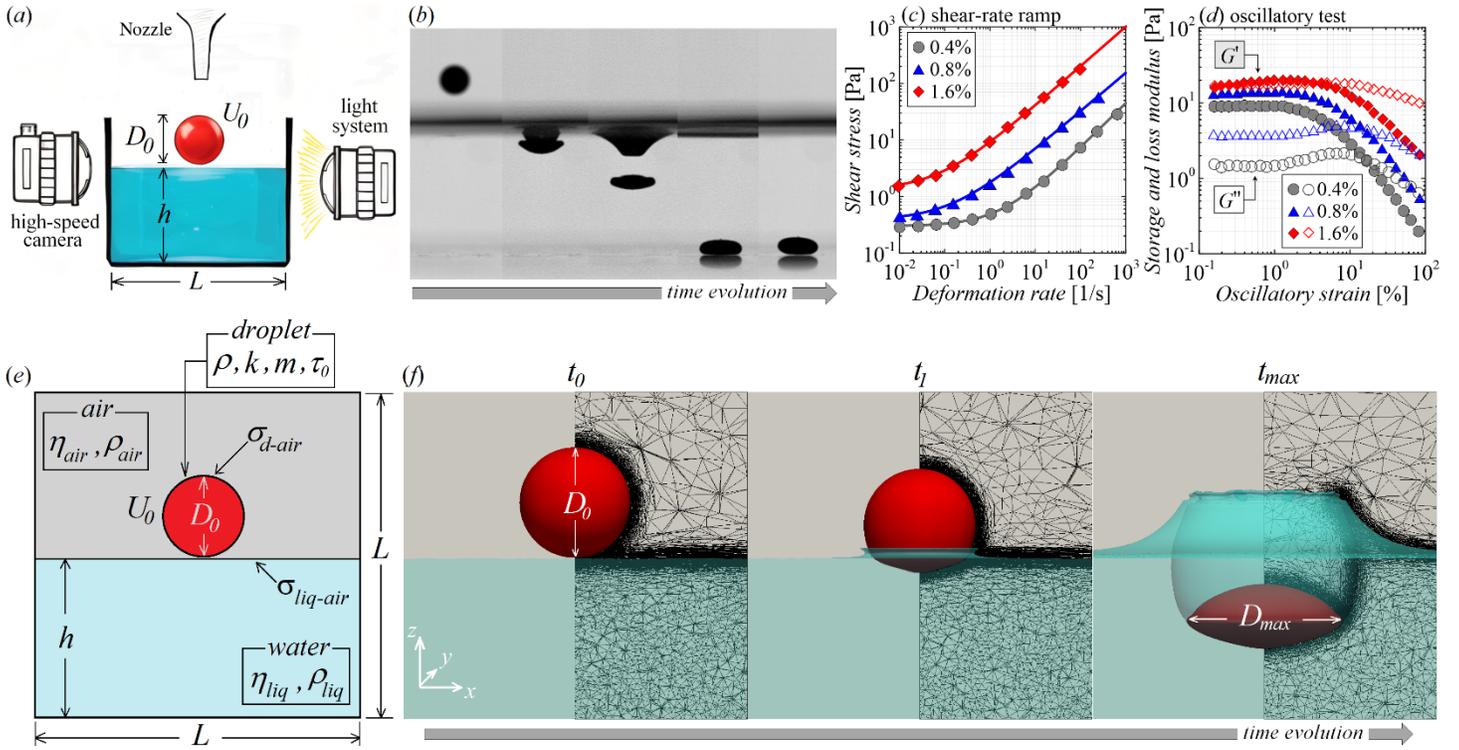

Figure 2. (a) Schematic illustration of the experimental setup. The non-Newtonian drop of diameter $D_0$ (in red) falls from a 400µm-diameter nozzle and impacts the interface between the reactive liquid (in blue) and the air (in white) at the velocity $U_0$. The penetration process is recorded by a high-speed camera with the aid of a light system. (b) Snapshots reconstruct the time evolution of a typical penetration event. (c-d) Rheological characterisation performed with a commercial rotative rheometer equipped with a cone-plate geometry for the three suspensions considered here: (c) shear stress against deformation rate; (d) storage and loss moduli against oscillatory strain. (e) Schematic representing the numerical scenario: a viscoplastic drop (in red) impacts the interface between the water (in blue) and the air (in grey). (f) Snapshots illustrating the time evolution of a typical impact-devoted 3D numerical simulation (the viscoplastic drop appears in red, the liquid is shown in blue, the air is illustrated in grey, and the mesh is depicted by the right-hand side black lines). A movie showing a numerical simulation is available on https://www.youtube.com/watch?v=srDNKYABsmY, in which $D_0$ = 2.5mm, $U_0$ = 1.4m/s, $k$ = 1.3Pa·s$^m$, $m$ = 0.68, $\tau_0$ = 0.38Pa, $\rho$ = 1200kg/m$^3$, $\rho_{liq}$ = 1000kg/m$^3$, $\eta_{liq} = 10^{-3}$Pa·s, $\rho_{air}$ = 1kg/m$^3$ and $\eta_{air} = 10^{-5}$Pa·s.

## 2. Problem statement, mixed experimental-numerical approach, and relevant dimensionless parameters

As previously mentioned, we present here an experimental and numerical study on the spreading of millimetric non-Newtonian drops undergoing chemically induced gelation while penetrating a reactive liquid, as illustrated in Fig. 1. These drops with an initial diameter $D_0$, density $\rho$, consistency $k$, yield stress $\tau_0$ and surface tension $\sigma$, fall under gravity $g$ from a nozzle and subsequently impact a liquid/air interface surface at velocity $U_0$ and time $t_0$. They then spread across the liquid/air interface, creating an air cavity until reaching a maximum spreading $D_{max}$ at time $t_{max}$. Surface tensions between the drop and the air, the drop and the liquid, and the liquid and the air are respectively $\sigma_{d-air}$, $\sigma_{d-liq}$, and $\sigma_{liq-air}$, while air density and viscosity are respectively represented by $\rho_{air}$ and $\eta_{air}$. The air cavity then closes under capillary and gravitational effects, compressing the drops and allowing them to penetrate the liquid of viscosity $\eta_{liq}$ and density $\rho_{liq}$ containing a gelling agent responsible for triggering gelation within the drops from the very beginning of the impact. Because $\rho$ is higher than $\rho_{liq}$, the drops move slowly towards the bottom of the container of dimensions L x L x L filled with reactive liquid, while the air cavity closes (L = 15cm; the volume of liquid is equal to L x L x h, where h = 5$D_0$). Finally, the drops tend to relax and assume a spherical shape due to the growth of the external elastic shell induced by gelation [1].

Since the relaxation process has been recently analysed, in the present work, we shed light on the drop spreading (black, red and blue squares in Fig.1) mainly focusing on the maximum spreading achieved by the drops $D_{max}$.

The penetration process is experimentally recorded using a Photon Fastcam Mini high-speed camera (10$^4$ frames per second) equipped with a 105mm macro lens and a focal doubler from Sigma, and with the aid of a LED backlight panel (50000cd/m² SLLUB White from Phlox; light system), as schematised in Fig.2(a). The time evolution of a typical penetration event is reconstructed by the snapshots in Fig.2(b). The initial diameter of the drops $D_0$ is approximately equal to 2.5mm, while the impact velocity $U_0$ (which is a function of the distance between the 400μm-diameter nozzle and the liquid/air interface) ranges from 0.5m/s to 2m/s.

Sodium alginate in 8% by volume zirconia-in-deionised-water drops are used at three alginate concentrations [Alg]: 0.4%, 0.8% and 1.6% by volume. The alginate is purchased from AGI with a molar mass of 250kg/mol and a polydispersity index of 1.7. The 150nm-radius zirconia particles are obtained from Saint-Gobain Zirpro (HanDan, China). We use zirconia particles to obtain a better contrast for imaging and to increase the density of the drops so that they sediment to the bottom of the liquid reservoir quickly after impacting the bath (Fig.1). Depletion interactions between the charged alginate (biopolymer) and the zirconia particles give rise to yield-stress fluids, i.e., materials that mechanically behave like an elastic solid at low-stress levels and a non-Newtonian liquid at stress levels above their yield stress $\tau_0$ with a shear-dependent viscosity $\eta$ [15-20]. The latter is commonly represented by the Herschel-Bulkley viscosity function (only valid to the flowing regions) according to which $\eta = k|\dot{\gamma}|^{m-1} + \tau_0/|\dot{\gamma}|$, where $k$ is the consistency index, $m$ is the flow index, and $|\dot{\gamma}|$ is the norm of the strain-rate tensor [21].

The drops are rheologically characterised using an ARES-G2 rheometer by TA Instruments equipped with a cone-plate geometry. Their relevant rheological properties are listed in Table 1. As indicated by the yield stress and the storage modulus $G'$, indeed these materials exhibit *elasto-viscoplastic* characteristics [15-17]. Some of them are illustrated in Figs.2(c) and 2(d) where the shear stress is displayed as a function of $|\dot{\gamma}|$, and the storage and the loss moduli, $G'$ (solid symbols) and $G''$ (open symbols) respectively, are plotted against the oscillatory strain at 10rad/s for the used drops: [Alg] = 0.4% (grey circles), [Alg] = 0.8% (blue triangles), and [Alg] = 1.6% (red diamonds). The stress curves in Figs.2(c) are fitted using the Herschel-Bulkley equation ($|\tau| = k|\dot{\gamma}|^m + \tau_0$, represented by the solid lines), which provides values of $k$, $m$ and $\tau_0$ associated with the non-Newtonian fluids before the impact. Furthermore, their relaxation time $\lambda$ can be estimated as $\lambda = [k/G'(|\tau| \to 0)]^{1/m}$ [22, 23]. By calculating the Weissenberg number $We = \lambda U_0/D_0 < 1$ (where $U_0/D_0$ is the flow characteristic strain rate), one can conclude that the initial drop elasticity tends to play a marginal role (at least for non-gelling events). Thus, the drops tend to behave as simple viscoplastic fluids [2, 24] throughout the liquid entry. In addition, since most surface tension measurement methods are corrupted by the fluid yield stress and elasticity, we follow previous works [22, 25] and take pure water surface tension as an upper bound for the real $\sigma_{d-air}$ of our drops, i.e., $\sigma_{d-air}$ < 0.072N/m. As a result, one can show that, for the flow cases analysed here, the ratio of the viscoplastic stress to the capillary stress associated with $\sigma_{d-air}$ (e.g., capillary number Ca) is always greater than 10, i.e., $Ca = \frac{k(U_0/D_0)^m + \tau_0}{\sigma_{d-air}/D_0} > 10$ [22]. In other words, the surface tension between the drop and the air $\sigma_{d-air}$ plays a marginal role when compared to viscoplastic effects. It is also essential to emphasise that $\sigma_{d-liq}$ is considered equal to zero, since the drops and the reactive liquids share the same solvent (deionised water).

Table 1. Rheological properties of the used sodium alginate in 8% by volume zirconia-in-deionised-water drops. Their density is approximatively equal to 1200kg/m³.

| [Alg] [%] | $k$[Pa·s$^m$] | $m$ | $\tau_0$[Pa] | $G'(|\tau| \to 0)$[Pa] |
|---|---|---|---|---|
| 0.4 | 0.19 | 0.79 | 0.3 | 9 |
| 0.8 | 1.3 | 0.69 | 0.38 | 12 |
| 1.6 | 8.02 | 0.72 | 1.3 | 19 |

The experimental drops impact a CaCl$_2$6H$_2$O-in-deionized-water solution (reactive liquid) at five different CaCl$_2$6H$_2$O concentrations (calcium chloride hexahydrate provided by Sigma-Aldrich): 0g/L, 1.58g/L, 5g/L, 15.8g/L, and 158g/L. Their density $\rho_{liq}$ and viscosity $\eta_{liq}$ are approximately equal to 10³kg/m³ and 10⁻³Pa·s, and $\sigma_{liq-air} \approx 0.072$N/m regardless the CaCl$_2$6H$_2$O concentration. The calcium mass diffusion coefficient within the drops K is approximately equal to 10⁻⁹m²/s [1-3, 9], which leads to a Peclet number $Pe = (D_0^2/K)/(D_0/U_0)$ lower than 10⁻¹² (note that $D_0^2/K$ is the characteristic calcium diffusion time within the drop, and $D_0/U_0$ is the flow characteristic time). Such a low Pe suggests that the spreading process would not be affected by gelation, a point that will be further investigated in Section 3. Lastly, $\rho_{air} = 1$kg/m³ and $\eta_{air} = 10^{-5}$Pa·s.

Based on the rationales presented above, one could imagine that, during the spreading process (early times), the yield-stress drops considered here behave as simple viscoplastic fluids not affected by capillary and gelation effects. Aiming to analyse these conjectures further, our experimental results are compared with three-dimensional (3D) numerical simulations devoted to the impact of simple viscoplastic drops on a liquid bath in absence of gelation, e.g., [CaCl$_2$6H$_2$O] = 0g/L (non-reactive

Email address for correspondence: * anselmo.soeiro pereira@minesparis.psl.eu | ** cecile.monteux@espci.fr

liquid). The numerical scenario is schematised in Fig.2(e) where the non-Newtonian drop is illustrated in red, the liquid is represented in blue, and the air appears in grey. Our computational method is based on a massively parallel finite element library [CIMLIB-CFD; 26-30] devoted to non-Newtonian multiphase flows [31-34]. More specifically, we apply the momentum conservation equation presented below to the solenoidal flow ($\nabla \cdot \boldsymbol{u} = 0$) described earlier (see Fig.2e):

$$\rho \left( \frac{\partial \boldsymbol{u}}{\partial t} + \boldsymbol{u} \cdot \nabla \boldsymbol{u} - \boldsymbol{g} \right) = -\nabla p + \nabla \cdot \boldsymbol{\tau} + \boldsymbol{f}_{st}, \quad (1)$$

where $\boldsymbol{u}$ is the velocity vector, $\nabla$ is the gradient operator, $\boldsymbol{g}$ is the gravity vector, $p$ is the pressure, $\boldsymbol{\tau}$ is the extra-stress tensor and $\boldsymbol{f}_{st}$ represents the force associated to the surface tension. The latter is defined as $\boldsymbol{f}_{st} = -\sigma \kappa \Phi \boldsymbol{n}$, where $\sigma$ is the surface tension, $\kappa$ is the curvature of the interfaces, $\Phi$ is the Dirac function and $\boldsymbol{n}$ is the normal vector at the interfaces. Furthermore, the extra-stress tensor is defined as $\boldsymbol{\tau} = \eta \dot{\boldsymbol{\gamma}}$, where $\dot{\boldsymbol{\gamma}}$ represents the rate-of-strain tensor $\dot{\boldsymbol{\gamma}} = (\nabla \boldsymbol{u} + \nabla \boldsymbol{u}^T)$. Our drops are described as Herschel-Bulkley fluids whose viscosity $\eta$ is given by

$$\eta = k|\dot{\boldsymbol{\gamma}}|^{m-1}\left(1 - e^{-|\dot{\boldsymbol{\gamma}}|/\dot{\gamma}_p}\right)^{1-m} + \frac{\tau_0}{|\dot{\boldsymbol{\gamma}}|}\left(1 - e^{-|\dot{\boldsymbol{\gamma}}|/\dot{\gamma}_p}\right). \quad (2)$$

As seen in Eq.2, this viscosity is coupled to the so-called Papanastasiou regularization (indicated in parenthesis) to avoid a numerical divergence when the norm of the rate-of-strain tensor $|\dot{\boldsymbol{\gamma}}|$ tends towards zero [31-34]. With this regularization term, the viscosity reaches a plateau when $|\dot{\boldsymbol{\gamma}}| < \dot{\gamma}_p$, where $\dot{\gamma}_p = 10^{-6}$s$^{-1}$. This plateau being enormous (between $10^5$ Pa·s and $10^{10}$ Pa·s), the drop behaves like a solid at the characteristic time scales of the spreading process (~ 1ms-10ms).

The time evolution of a typical three-phase numerical simulation is illustrated by the snapshots in Fig.2(f). The numerical methods used here are based on a *Variational Multiscale Method* (*VMS*) coupled with an anisotropic mesh adaptation method [26-34]. The meshes are composed of approximately $5 \times 10^6$ elements whose minimum size is 1μm [see the black lines in Fig.2f]. The Courant–Friedrichs–Lewy (CFL) condition is lower than 0.1 for all numerical simulations. The evolution of the interfaces over time is described using a Level-Set function [26-35]. Fluid properties are interpolated through the level-set functions whose value varies from positive within the drop and the liquid bath to negative in the air (and zero at the interfaces between the different fluid phases; supplemental details concerning the interface capturing and the level-set functions are available in Refs. 30 and 35]. Initial and boundary conditions for the flow equations are, respectively, initial vertical impact velocity $U_0$ within the drop, zero normal stress in the upper wall of the computational domain, and no-slip condition between the fluids and the other walls.

The fluid properties considered in the simulations are similar to those measured experimentally for all three fluid phases (impacting drop, impacted liquid, and air). The impact velocity ranges from 0.5m/s to 2m/s, and the initial drop diameter is fixed at 2.5mm. A movie showing a typical numerical simulation is available on https://www.youtube.com/watch?v=srDNKYABsmY, in which $D_0$ = 2.5mm, $U_0$ = 1.4m/s, $k$ = 1.3Pa·s$^m$, $m$ = 0.68, $\tau_0$ = 0.38Pa, $\rho$ = 1200kg/m$^3$, $\rho_{liq}$ = 1000kg/m$^3$, $\eta_{liq}$ = $10^{-3}$Pa·s, $\rho_{air}$ = 1kg/m$^3$ and $\eta_{air}$ = $10^{-5}$Pa·s (note the generation of the bowl, which is equally observed in the correspondent experiment shown in Fig.3a).

The dimensionless numbers dominating the problem, with respect to a simple viscoplastic drop, are denoted as $\Pi_i$ and obtained from the Buckingham-$\Pi$ theorem whose variables are $D_0, U_0, \rho, k, \tau_0, \rho_{liq}, \eta_{liq}, \sigma_{liq-air}, \rho_{liq}, \eta_{liq}$, and $g$ (downwards component of $\boldsymbol{g}$), while the fundamental units are mass [kg], distance [m] and time [s]. This results in eight important dimensionless quantities associated with the considered three-phase fluid problem:

$$\Pi_1 = \frac{k(U_0/D_0)^m}{\rho U_0^2}, \quad (3)$$

$$\Pi_2 = \frac{\tau_0}{\rho U_0^2}, \quad (4)$$

$$\Pi_3 = \frac{\rho g D_0}{\rho U_0^2}, \quad (5)$$

$$\Pi_4 = \frac{\sigma_{liq-air}/D_0}{\rho U_0^2}, \quad (6)$$

$$\Pi_5 = \frac{\eta_{liq}(U_0/D_0)}{\rho U_0^2}, \quad (7)$$

$$\Pi_6 = \frac{\eta_{air}(U_0/D_0)}{\rho U_0^2}, \quad (8)$$

$$\Pi_7 = \frac{\rho_{liq}}{\rho}, \quad (9)$$

$$\Pi_8 = \frac{\rho_{air}}{\rho}, \quad (10)$$

where $\Pi_1 = 1/\text{Re}_m$ ($\text{Re}_m$ is the power-law-based Reynolds number), $\Pi_2 = \text{Pl}$ (Pl denotes the Plastic number), $\Pi_3 = 1/\text{Fr}$ (Fr indicates the Froude number), $\Pi_4 = 1/\text{We}$ (We is the Weber number), $\Pi_5 = 1/\text{Re}_{liq}$ ($\text{Re}_{liq}$ is the liquid-based Reynolds number), $\Pi_6 = 1/\text{Re}_{air}$ ($\text{Re}_{air}$ is the air-based Reynolds number), and $\Pi_7 = \rho_{liq}/\rho$ and $\Pi_8 = \rho_{air}/\rho$ are density ratios. Our

Email address for correspondence: * anselmo.soeiro pereira@minesparis.psl.eu | ** cecile.monteux@espci.fr

impacting objects being millimetric and presenting a low yield stress (≤ 1.3Pa, as indicated in Fig.2c and Table 1), $\rho g D_0$ and $\tau_0$ become negligible compared to the inertial stress $\rho U_0^2$, e.g., $\Pi_2 < 10^{-2}$ and $\Pi_3 < 10^{-1}$. Moreover, $\Pi_4 < 10^{-1}$, $\Pi_5 < 10^{-6}$, and $\Pi_6 < 10^{-8}$, while $\Pi_7$ and $\Pi_8$ are kept respectively fixed at 0.8333 and $10^{-3}$. As a result, at first glance, the spreading process seems to be dominated by only one dimensionless number:

$$\text{Re}_m = \frac{\rho U_0^2}{k(U_0/D_0)^m}. \qquad (11)$$

In other words, the early time spreading of the impacting non-Newtonian drops considered here would be purely driven by a competition between inertial and viscous stresses and, thus, not affected by gelation, as mentioned previously. This conjecture is further analysed through experiments and numerical simulations in the following Section.

## 3. Results and discussions

Figure 3 illustrates the penetration of 8% by volume zirconia-in-deionised-water drops with 0.8% by volume of sodium alginate within a liquid bath at five different $CaCl_2 6H_2O$ concentrations: 0g/L (Fig.3a), 1.58g/L (Fig.3b), 5g/L (Fig.3c), 15.8g/L (Fig.3d), and 158g/L (Fig.3e). Each subfigure is composed of ten snapshots detailing the penetration process, from the drop impact on the liquid/air interface at 1.4m/s, until the drop sedimentation and eventual shape relaxation. It is essential to observe that the shapes of the impacting objects are clearly different at $t_{max}$ (≈ 9ms). More specifically, as the concentration of calcium ions increases, the shape of the drops at $t_{max}$ evolves from a bowl (Figs.3a and 3b) to an ellipsoid (Figs.3d and 3e), which reveals that gelation plays a non-negligible role on the drop penetration dynamics from its very beginning. In addition, after sedimentation, gelation induces shape relaxation because of the generation of a growing elastic shell. The relaxation towards a spherical form appears as an increasing function of the calcium ions concentration, a topic deeply discussed by Godefroid and co-authors [1]. On the other hand, in the absence of $CaCl_2 6H_2O$, the drop simply dissolves with time, as displayed in Fig.3a.

Supplemental early-time gelation effects are shown in Figs.4(a) and 4(b) where the drop diameter $D(t)$ (made dimensionless by $D_0$) is displayed as a function of time $t$. We consider 0.8% by volume alginate drops impacting at 1.4m/s (Fig.4a) and 2m/s (Fig.4b) on liquid baths at four different $CaCl_2 6H_2O$ concentrations represented by the

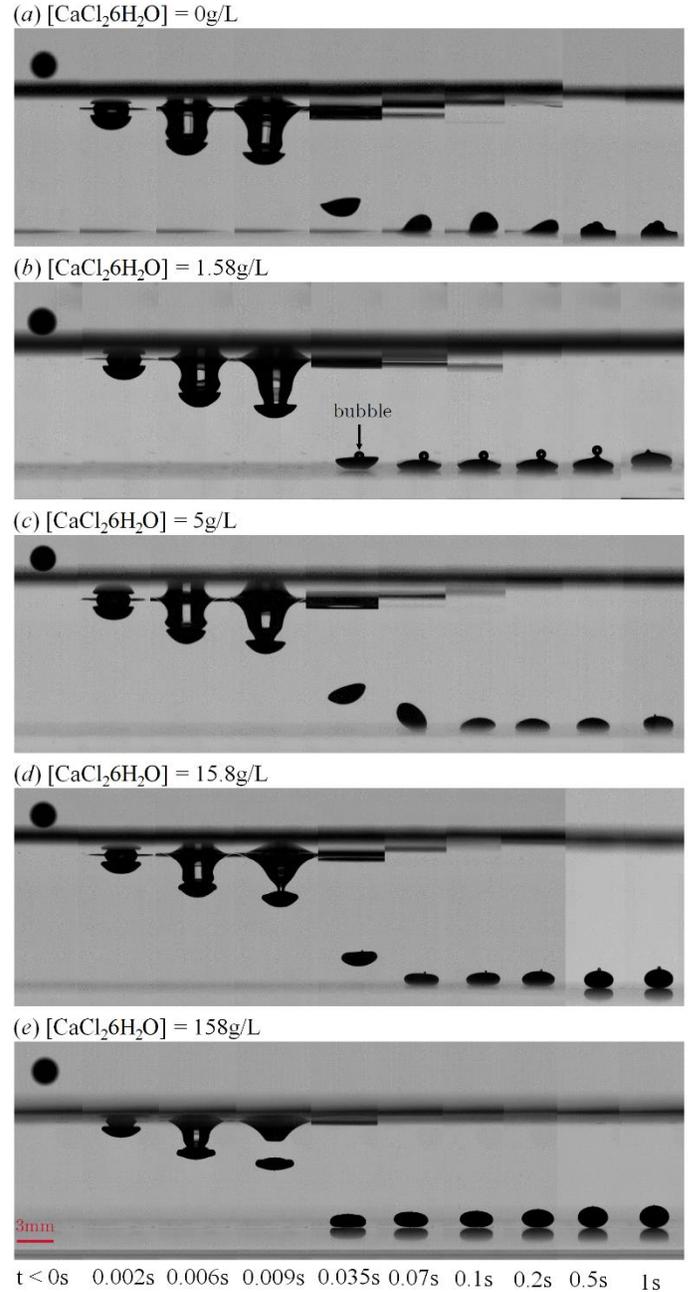

**Figure 3.** Experimental snapshots illustrating the penetration of an 8% by volume zirconia-in-deionised-water drop with 0.8% by volume of sodium alginate (biopolymer) into a liquid bath at five $CaCl_2 6H_2O$ concentrations: (a) 0g/L; (b) 1.58g/L; (c) 5g/L; (d) 15.8g/L; (e) 158g/L. The falling 2.5mm-diameter non-Newtonian drops impact the liquid/air interface at 1.4m/s.

symbols: 0g/L (grey circles), 5g/L (blue triangles), 15.8g/L (red diamonds), and 158g/L (green rectangles). Regardless of the calcium ions concentration, gelation effects on $D_{max}$ seem marginal at $U_0$ = 1.4m/s, as previously indicated by Fig.3. However, a contrasting scenario emerges by increasing $U_0$ up to 2m/s, drop impact velocity at which $D_{max}$ clearly becomes a decreasing function of the

Godefroid et al. - Gelation effects on the spreading of non-Newtonian drops impacting a reactive liquid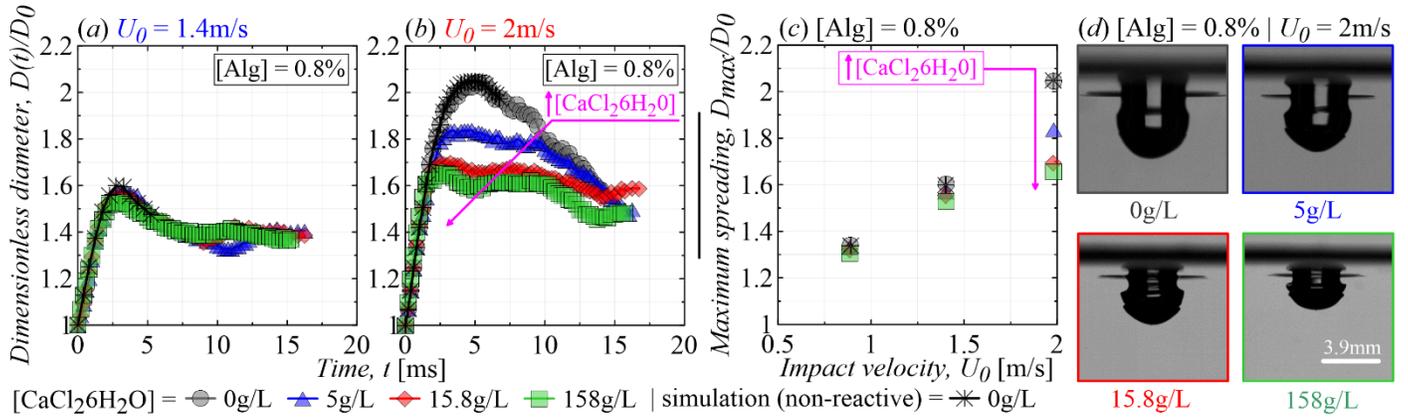

**Figure 4.** (a-b) Dimensionless diameter $D(t)/D_0$ as a function of the dimensionless time $t/t_{max}$ for 8% by volume zirconia-in-deionised-water drops with 0.8% by volume of sodium alginate impacting at 1.4m/s (a) and 2m/s (b) reactive liquid baths at four $CaCl_2 6H_2O$ concentrations: 0g/L (grey circles); 5g/L (blue triangles); 15.8g/L (red diamonds); and 158g/L (green rectangles). (c) Dimensionless maximum spreading $D_{max}/D_0$ as a function of the impact velocity $U_0$ for 8% by volume zirconia-in-deionised-water drops with 0.8% by volume of sodium alginate impacting reactive baths at 0.89m/s, 1.4m/s, and 2m/s (snapshots at $t_{max}$ are displayed in d). Numerical simulations performed for non-gelling impact scenarios (black asterisks in a, b and c) are compared with the experimental curves at $[CaCl_2 6H_2O]$ = 0g/L (grey circles).

$CaCl_2 6H_2O$ concentration, as pointed out by the magenta arrow. Nevertheless, it is worth noting that a $CaCl_2 6H_2O$ saturation level tends to be achieved, from which $D(t)/D_0$ reaches a minimum level for a fixed $U_0$ and alginate concentration. This saturation tendency is observed by comparing the red diamonds (15.8g/L) with the green rectangles (158g/L) in Fig.4(b).

The correlation between the impact velocity and the calcium ions concentration is further underlined by Fig.4(c) in which the dimensionless maximum spreading $D_{max}/D_0$ is plotted as a function of $U_0$ for 0.8% by volume alginate-in-deionised-water drops penetrating within liquid baths at different $CaCl_2 6H_2O$ concentrations. For the considered alginate-based drops, gelation effects on $D_{max}$ become relevant only at $U_0 > 1.4$m/s. Such a correlation can be justified by the impact velocity-induced increase in contact area between the drop and the liquid at the very beginning of the penetration process, which favours the $CaCl_2 6H_2O$ diffusion into the drop and, consequently, the generation of an elastic shell able to constrain the spreading at early times. Consistently, this effect becomes more pronounced with an increasing calcium ions concentration until the saturation level is achieved, as illustrated by the snapshots in Fig.4(d) taken at $t_{max}$ for a 0.8% by volume alginate-in-deionised-water drop impacting at 2m/s on liquid baths with a $CaCl_2 6H_2O$ concentration ranging from 0g/L to 158g/L (i.e., snapshots showing $D_{max}$ at $t_{max}$). It is also interesting to note that, like $D_{max}$, $t_{max}$ is a decreasing function of $[CaCl_2]$, as highlighted by Fig.4(b). Therefore, a lower penetration level is observed at 158g/L in Fig.4(d).

Numerical simulations performed for non-gelling impact scenarios (black asterisks in Figs.4a, 4b and 4c) are compared with the experimental curves at $[CaCl_2 6H_2O]$ = 0g/L (grey circles). The good agreement between experiments and numerical simulations gives us confidence to explore numerically the physical mechanism driving the spreading dynamics in light of kinematic and energy transfer analyses. Hence, first, we show in Figs.5(a) and 5(b) the numerical contours of the norm of the instantaneous radial velocity $|u_y|$, made dimensionless by the instantaneous maximum radial velocity $u_{y,max}$, on the centre x-z plane at $t = 0.25 t_{max}$ (Fig.5a) and $t = 0.5 t_{max}$ (Fig.5b). The velocity field strongly suggests that, during the spreading process, the drop undergoes bi-axial extension, which is corroborated by Fig.5(c) where the dimensionless volume-averaged deformation rate terms $\dot{\gamma}_e^*$ (magenta stars) and $\dot{\gamma}_s^*$ (cyan crosses) are plotted as a function of the dimensionless time $t/t_{max}$. These terms are defined as

$$\dot{\gamma}_e^* = \frac{1}{V}\int_V \frac{1}{2|\dot{\gamma}|^2}\left(\dot{\gamma}_{xx}^2 + \dot{\gamma}_{yy}^2 + \dot{\gamma}_{zz}^2\right) dV, \qquad (12)$$

and

$$\dot{\gamma}_s^* = 1 - \dot{\gamma}_e^*, \qquad (13)$$

where $V$ is the volume of the drop, $\dot{\gamma}_{xx} = 2\partial u_x/\partial x$, $\dot{\gamma}_{yy} = 2\partial u_y/\partial y$, and $\dot{\gamma}_{zz} = 2\partial u_z/\partial z$. Therefore, $\dot{\gamma}_e^*$ is exclusively linked with extension/compression, while $\dot{\gamma}_s^*$ is shear based. Clearly, $\dot{\gamma}_e^*$ dominates $\dot{\gamma}$ throughout the drop spreading. Lastly, numerical analyses of the time evolution of the drop's kinetic energy and dissipation are performed in Fig.5(*c*). These terms are calculated as

$$KE(t) = \int_V \frac{\rho|\boldsymbol{u}(t)|^2}{2}\, dV, \quad (14)$$

and

$$W(t) = \int_t \int_V \frac{k}{m+1}|\dot{\boldsymbol{\gamma}}(t)|^{m+1} + \tau_0|\dot{\boldsymbol{\gamma}}(t)|\, dVdt. \quad (15)$$

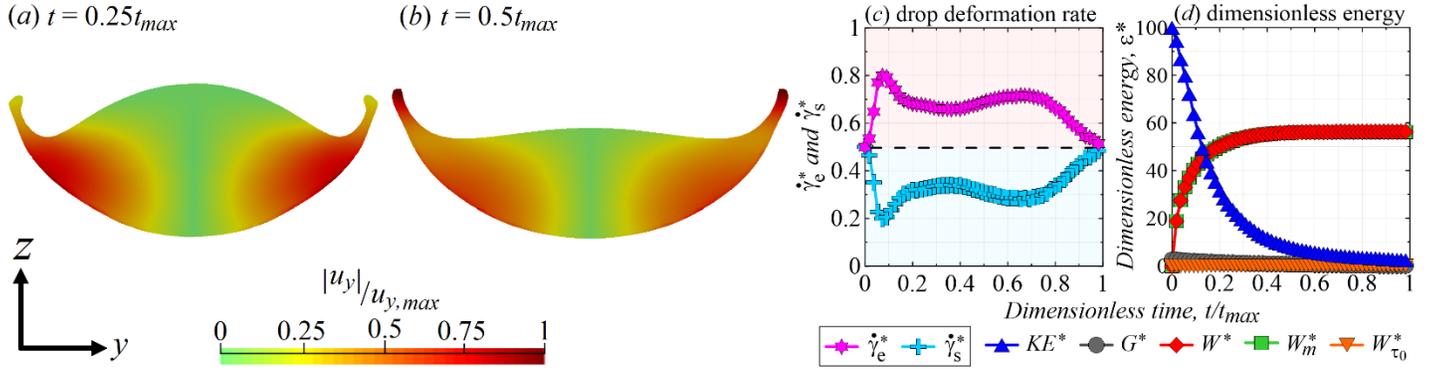

**Figure 5.** 3D numerical simulation: $U_0$ = 1.4m/s, $k$ = 1.3Pa·$s^m$, $m$ = 0.68, and $\tau_0$ = 0.38Pa, e.g., [Alg] = 0.8%. (*a-b*) Numerical contours of the dimensionless radial velocity $|u_y|/u_{y,max}$ on the centre *x-z* plane at $t = 0.25t_{max}$ (*a*) and $t = 0.5t_{max}$ (*b*). Note the generation of the bowl, which is equally observed at the correspondent experiment shown in Fig.3(*a*). (*c-d*) Dimensionless volume-averaged deformation rate terms associated with extension/compression ($\dot{\gamma}_e^*$; pink stars in the ping region), and shear ($\dot{\gamma}_s^*$; cyan crosses in the blue region). (*d*) Dimensionless energy terms as a function of the dimensionless time $t/t_{max}$.

The dissipated energy is made up of its power-law contribution

$$W_m = \int_t \int_V \frac{k}{m+1}|\dot{\boldsymbol{\gamma}}|^{m+1}dVdt \quad (16)$$

and yield-stress-based contribution

$$W_{\tau_0} = \int_t \int_V \tau_0|\dot{\boldsymbol{\gamma}}|\, dVdt. \quad (17)$$

Each energy term is made dimensionless by the initial kinetic energy $KE_0 = KE(t=0)$: $KE^* = KE/KE_0$ x 100[%] (blue triangles), $W^* = W/KE_0$ x 100[%] (red diamonds), $W_m^* = W_m/KE_0$ x 100[%] (green squares), and $W_{\tau_0}^* = W_{\tau_0}/KE_0$ x 100[%] (orange inverted triangles). They are generically denoted as $\varepsilon^*$ and finally plotted against the dimensionless time $t/t_{max}$, revealing that the dissipation within the drop is comparable with the drop's kinetic energy [$W^*(t = t_{max}) \approx 0.6KE^*(t=0)$; the rest of the initial kinetic energy is transferred to the liquid, a result not shown for brevity]. Hence, together with the velocity contours and the deformation rate curves illustrated in Fig.5(*a*), 5(*b*), and 5(*c*), the energy budget indicates that the spreading process is primarily dissipative and bi-axial extensional. Furthermore, we

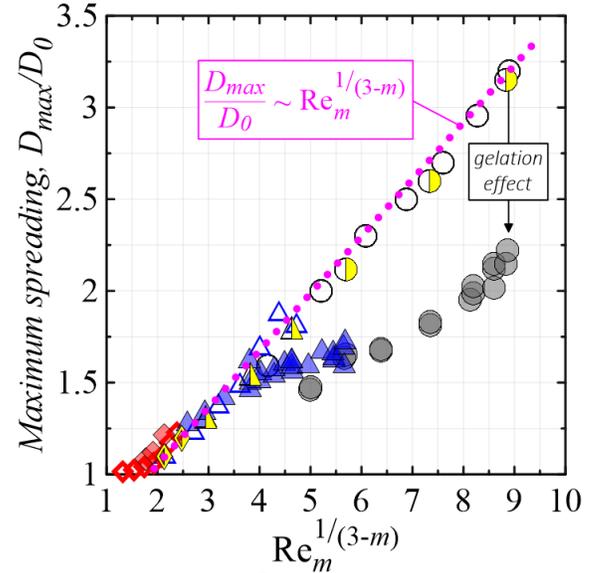

exp. [CaCl$_2$] = 158g/L [● 0.4%  ▲ 0.8%  ◆ 1.6%]
exp. [CaCl$_2$] = 0g/L [◐ 0.4%  ▲ 0.8%  ◆ 1.6%]
sim. [CaCl$_2$] = 0g/L [○ 0.4%  △ 0.8%  ◇ 1.6%]

**Figure 6.** Drop maximum spreading $D_{max}$, made dimensionless by $D_0$, as a function of $\text{Re}_m^{1/(3-m)}$. Experimental results at [CaCl$_2$6H$_2$O] = 0g/L (half-filled symbols; non-gelling) and 158g/L (solid symbols) are plotted with numerical ones (open symbols) associated with non-gelling impact events. Drops at three alginate concentrations are considered: 0.4% (grey circles), 0.8%

observe in Fig.5(*c*) that the dissipation within the drop is essentially dominated by $W_m^*$ (i.e., $W_m^* \gg W_{\tau_0}^*$; $W_{\tau_0}^* \approx 0$), and $G^* < 2.5\%$. In other words, yield-stress and gravitational effects are negligible, as anticipated in Section 2. These tendencies are confirmed by all simulations presented in this work.

Based on the results highlighted by Fig.5, we propose a scaling law for $D_{max}$ by considering that, during the gelation-free spreading of the drop, the impacting kinetic energy $[\sim \rho U_0^2 D_0^3]$ is mainly dissipated throughout a bi-axial extensional flow with negligible yield-stress effects $[\sim k \dot{\gamma}_c^m D_{max}^3$, where $\dot{\gamma}_c$ is the characteristic deformation rate defined as $\dot{\gamma}_c \sim U_0/D_{max}$ due to its bi-axial extensional nature; 35]. By equating this energy terms, we find that

$$\frac{D_{max}}{D_0} \sim \text{Re}_m^{1/3-m}. \quad (18)$$

Note that for Newtonian fluids ($m = 1$), $D_{max}/D_0$ scales with $\text{Re}_m^{1/2}$, a previously reported inertio-viscous scaling law linked with the spreading of viscous drops under free-slip and inverse Leidenfrost conditions [e.g., bi-axial extensional spreading; 29]. The validity of Eq.18 is also corroborated by Fig.6 where $D_{max}/D_0$ is displayed against $\text{Re}_m^{1/(3-m)}$ for both numerical cases (open symbols) associated with non-gelling impact events, and experimental cases at [CaCl$_2$6H$_2$O] = 0g/L (non-gelling events; half-filled yellow symbols) and 158g/L (gelling events; solid symbols). Drops at three alginate concentrations are considered: 0.4% (circles), 0.8% (triangles), and 1.6% (diamonds) by volume. The numerical cases are perfectly fitted by Eq.18 (represented by the pink dotted line) for the whole range of Reynolds number considered here, as well as the experimental non-gelling events (yellow symbols). Moreover, experimental and numerical results collapse for $\text{Re}_m^{1/(3-m)} < 3.5$, which is in line with the negligible gelation effects on the drop's maximum spreading at low impact velocities stressed by Fig.3 and Fig.4(a). However, as $\text{Re}_m^{1/(3-m)}$ becomes more pronounced, numerical and experimental gelling results diverge from each other. As discussed previously, gelation effects on $D_{max}$ emerge from the Reynolds-number-induced increase in the drop/liquid contact area at early times, which favours the calcium ions diffusion into the drops, ultimately generating an elastic shell able to oppose the spreading from the very beginning of the liquid entry, despite the high Peclet number associated with the impact events.

In short, even at very low $\text{Pe}$, gelation can lead to a drop spreading regime which differs dramatically from the bi-axial extensional-based inertio-viscous one observed for non-gelling situations under negligible yield-stress effects $[D_{max}/D_0 \sim \text{Re}_m^{1/(3-m)}]$. Nevertheless, the physicochemical hydrodynamics driving this new regime remains to be deeper explored in future works.

## 4. Concluding remarks and perspectives

We have presented a mixed experimental and numerical study highlighting relevant gelation effects on the early-time spreading (< 10ms) of millimetric non-Newtonian drops of biopolymer (sodium alginate) and particle (zirconia) suspensions impacting a Newtonian liquid containing reactive compounds (CaCl$_2$6H$_2$O). Our analyses have been conducted by considering a range of biopolymer and gelling compound concentrations and drop impact velocities.

In the absence of gelling agents (non-reactive events), the considered non-Newtonian drops behave like simple Herschel-Bulkley fluids with negligible yield stress. A competition between inertial stresses and bi-axial extensional viscous stresses drives their spreading. Therefore, their maximum spreading $D_{max}$ scales with $\text{Re}_m^{1/(3-m)}$, where $\text{Re}_m$ is the power-law-based Reynolds number $[\text{Re}_m = \rho U_0^2 / k(U_0/D_0)^m]$.

A new spreading regime emerges when gelling agents are added to the liquid bath, leading to dramatically lower $D_{max}$ values (i.e., gelling impact events do not follow the inertio-viscous scaling mentioned above). The increase of $\text{Re}_m$ favours such effects. More specifically, as $\text{Re}_m$ becomes more pronounced, the drop spreading upon the reactive liquid/air interface increases, which leads to the enhancement of the contact area between the drop and the liquid, and consequently to a faster CaCl$_2$6H$_2$O diffusion into the drop. This diffusion triggers the generation of an elastic shell in the outer part of the impact object, which can constrain the drop spreading at early times and ultimately decrease $D_{max}$ compared to the corresponding non-reactive impact event. Naturally, this effect is equally favoured by an increasing gelling agent concentration until a saturation level is achieved.

Finally, in terms of perspectives, since gelling drops are present in numerous applications (as underlined in Section 1), it is important to consider in near-future works supplemental analyses allowing to translate the gelation-induced rheological evolution of the drops into scaling laws predicting $D_{max}$ as a function of the important parameters of the problem (impact velocity, drop initial morphology, polymer concentration, gelling agent concentration etc.). Extending the results to yield-stress-affected impact events would also be greatly appreciated.

## Availability of materials and data

The datasets generated during and/or analysed during the current study are available from the corresponding author on reasonable request.

## Conflicts of interest

The authors declare no conflicts of interest.

## Acknowledgements

We would like to thank D. Edith Peuvrel-Disdier (CNRS and Mines Paris - PSL, France), D. Kindness Isukwem (The University of British Columbia, Canada) and D. Etienne Barthel (CNRS and ESPCI Paris - PSL) for their valuable comments and suggestions, and Mrs. Camila Borgo for her great help with the art in Fig.1. We are grateful for the financial support granted to us by Saint-Gobain for the PhD thesis of JG. AP acknowledges the UCA$^{JEDI}$ program (IdEx of Université Côte d'Azur), the PSL Research University under the program 'Investissements d'Avenir' launched by the French Government and implemented by the French National Research Agency (ANR) with the reference ANR-10-IDEX-0001-02 PSL, and the ANR for supporting the INNpact project (ANR-21-CE06-0036) under the 'Jeunes chercheuses et jeunes chercheurs' program.

Email address for correspondence: * anselmo.soeiro pereira@minesparis.psl.eu | ** cecile.monteux@espci.fr

Email address for correspondence: * anselmo.soeiro pereira@minesparis.psl.eu | ** cecile.monteux@espci.fr